\documentclass[twocolumn,aps,floats,superscriptaddress,showpacs]{revtex4}

\usepackage{amsmath}
\usepackage{epsfig}
\usepackage{graphicx}
\usepackage{dcolumn}
\usepackage{bm}

\newcommand{\be}{\begin{equation}}
\newcommand{\ee}{\end{equation}}
\newcommand{\bea}{ \begin{eqnarray}}
\newcommand{\eea}{ \end{eqnarray}}
\newcommand{\Eq}[1]{Eq.\,(\ref{#1})}
\newcommand{\Fig}[1]{Fig.\,\ref{Fig#1}}

\renewcommand{\Im}{\mbox{\sl Im\,}}
\renewcommand{\Re}{\mbox{\sl Re\,}}
\renewcommand{\hom}{\hbar \omega}
\renewcommand{\r}{{\bf r}}
\newcommand{\R}{{\bf R}}
\renewcommand{\k}{{\bf k}}
\newcommand{\kt}{\tilde{\bf k}}
\newcommand{\q}{{\bf q}}
\newcommand{\G}{{\cal G}}
\newcommand{\DeltaR}{\Delta_{\r}}
\newcommand{\hz}{\hbar z}
\newcommand{\Sz}{\Sigma(z)}
\newcommand{\Sk}{\Sigma_{\k}(z)}
\newcommand{\Skl}{\Sigma_{\k'}(z)}
\newcommand{\ek}{\epsilon_\k}
\newcommand{\ekl}{\epsilon_{\k'}}
\newcommand{\gkk}{g_{\k-\k'}}
\newcommand{\gk}{g_{\k}}

\begin{document}

\title{Coherent potential approximation for spatially correlated disorder}
\author{Roland Zimmermann}\email{zim@physik.hu-berlin.de}
\affiliation{Institut f\"ur Physik der Humboldt-Universit\"at zu
Berlin, Newtonstr. 15, D-12489 Berlin, Germany}
\author{Christoph Schindler$^{1,}$}
\affiliation{Walter Schottky Institut, Technische Universit\"at
M\"unchen, D-85748 Garching, Germany}

\date{\today}

\begin{abstract}
The coherent-potential approximation (CPA) is extended to describe
satisfactorily the motion of particles in a random potential which
is spatially correlated and smoothly varying. In contrast to
existing cluster-CPA methods, the present scheme preserves the
simplicity of the conventional CPA  but leads to a momentum and
frequency-dependent self-energy. Its accuracy is checked by a
comparison with the exact moments of the Green's function, and with
the spectral function from numerical simulations. The scheme is
applied to excitonic absorption spectra in different spatial
dimensions.
\end{abstract}

\pacs{71.35.Cc, 72.80.Ng, 78.40.Pg, 78.20.Bh}

\maketitle

\section{Introduction} \label{Introduction}

Disorder is a pertinent feature in many solid state systems, and has
been a topic of interest and intense research for many decades. The
prototype system is a binary alloy where the chemical constituents
are randomly (and spatially uncorrelated) placed on the sites of a
regular lattice. Adding next neighbor hopping for the electronic
band under consideration, and leaving out any additional
interactions, we are left with the discrete Anderson model which has
been widely studied, mostly in view of calculating the electronic
density of states.

A break through in the theoretical treatment was the invention of
the coherent-potential approximation (CPA), initially proposed by
Soven \cite{Soven} and further developed by Velicky, Kirkpatrick,
and Ehrenreich \cite{VKE,Velicky}. In a nutshell, the system is
replaced by a perfect crystal (effective medium) with a complex self
energy instead of the disorder potential, whose value is fixed by
demanding that the electron at the central site is not scattered off
the surrounding medium (on the average). The standard (single-site)
version of the CPA is characterized by a momentum-independent self
energy, and  leaves out finer details in the density of states. It
has been extended by embedding not only a single site but a larger
cluster into the effective medium. Consequently, in this embedded
cluster CPA \cite{Cluster}, a self-energy matrix instead of a single
element has to be determined self-consistently. This can work
satisfactorily for one spatial dimension \cite{Eggarter}, but apart
from increasing the numerical work, it is prone to numerical
instabilities  and shows sometimes an incorrect analytic behavior.
Such problems have been reported by Mills and co-workers
\cite{Mills} when introducing the "travelling cluster method" for
the bulk case.

Not unexpected, a single-site approximation will fail completely if
already the underlying disorder has a spatial correlation. This
happens for an alloy with clustering in the chemical composition.
For bulk semiconductor material with impurities, the finite range of
the impurity potential itself gives rise to a spatial correlation,
independent on the placement of the impurity atoms. Among the papers
which have applied and refined cluster-CPA methods for correlated
disorder, we quote Ref.~\cite{Mookerjee} where further references on
earlier work can be found. Systematic improvements have been
achieved by Jarrell applying his dynamical cluster approximation
(DCA) to disorder problems \cite{Jarrell}. This approach has been
later adopted to correlated disorder as well \cite{Maier}. However,
in order to preserve the correct analytic properties in the DCA, the
momentum dependence of the self-energy had to be coarse grained
according to the cluster size used (see the critical discussion of
Rowlands \cite{Rowlands}). A cluster CPA with only one complex
function of frequency to be determined self-consistently has been
derived recently in Ref.~\cite{Laad}, but was applied to
next-neighbor correlations only.

Our main motivation to deal with correlated disorder came from
excitons (Coulomb-bound electron-hole pairs) where the averaging
with the ground-state wave function produces a correlated potential
quite naturally. The reader who is not so much interested in exciton
physics may skip the corresponding Sec.~\ref{Excitons} and proceed
immediately with the formulation of the problem in terms of the
continuous Anderson model (Sec.~\ref{Formulation}). The
coherent-potential approximation is extended to a (smoothly)
correlated potential landscape in Sec.~\ref{CorrelatedCPA},
introducing a momentum and frequency-dependent self-energy.  The
resulting scheme preserves the simplicity of the standard CPA in so
far as for each momentum and frequency, a single root searching in
the complex plane is sufficient. In Sec.~\ref{Moments}, the exact
moments of the spectral function are compared with those of the
correlated CPA. Numerical results for the spectral function are
presented in Sec.~\ref{Results} and compared to numerically exact
simulation results. For the latter, the Kernel polynomial method
(KPM) \cite{Alvermann} is applied which rests upon an expansion of
the Hamiltonian into Chebyshev polynomials. In Appendix A it is
shown that the present method guarantees the proper analytic
behavior (Herglotz properties) of the self-energy in the
complex-frequency plane. Explicit expressions for several functions
needed in the correlated CPA are listed in the Appendix B while in
Appendix C results for the exact moments are collected.

\section{Application: Excitons in disordered semiconductors}
\label{Excitons}

Pioneering work on excitons and disorder has been done by
Baranovskii and Efros \cite{Baranovski} emphasizing the distinct
role of the relative motion of electron and hole within the exciton
in contrast to the center-of-mass (cm) motion of the exciton as a
whole. Both motions average over the underlying disorder and lead to
a double-step "motional narrowing" which gives rise to an
inhomogeneous exciton line width well below the width of the
band-edge fluctuations. In the first (relative motion) step one
needs a proper definition of the "excitonic volume" which is related
to the integral over the fourth power of the relative wave function
\cite{Zim89,LeeBajaj}. This has been refined later to take into
account the magnetic-field dependence of the relative exciton motion
\cite{Lyo}. The implementation of the second (cm motion) step was
more complicated. An early attempt \cite{Kanehisa} was using a
CPA-like treatment of electrons and holes separately but has
decoupled the formation of excitons from the disorder influence
following Ref.~\cite{Velicky}.

If the exciton is well localized on a single lattice site (Frenkel
exciton), only the cm motion has to be considered. This is the realm
of excitations in tight-binding chains where disorder treatments
have a long history. An early comparison between simulation and CPA
for uncorrelated potentials has been presented by Huber and Ching
\cite{Huber}. The corresponding motional narrowing has been
discussed widely based on simulations \cite{Reineker}. Correlated
disorder of a very specific form (dimers of identical energy) has
been treated in Ref.~\cite{Adame} via numerical averaging over
disorder realizations. In Ref.~\cite{Bakalis}, Frenkel excitons in
silver-halide systems have been investigated using the standard CPA.
Assuming still uncorrelated disorder, the model has been generalized
to potential distributions more complicated than simple Gaussian.

In the case of moderate disorder (the inhomogeneous exciton line
width being small compared with its binding energy), the
ground-state exciton wave function for relative motion may be
treated as unaffected by disorder. Then, the task reduces to solve
an effective one-particle problem for the cm motion. Still, the
relative motion was reducing the fluctuations of the effective cm
potential (first step motional narrowing). However, much more
important is that the resulting cm potential is now spatially
correlated at least over distances comparable to the exciton Bohr
radius -- even if the underlying band-edge fluctuations were not
correlated at all. This was our main stimulus to treat the exciton
cm motion in a correlated potential.

A straightforward way to do so are numerical simulations: For a
given potential, the Schr\"odinger equation is solved for
eigenvalues and eigenfunctions, wherefrom the spectral function can
be extracted easily. Note that at zero cm momentum, the spectral
function agrees with the absorption line shape of the
inhomogeneously broadened exciton. A lot more effective is to solve
the time-dependent Schr\"odinger equation for the Green's function
which gives upon time Fourier transformation the spectral function
directly \cite{Huber,GlutschTime}. Mandatory, however, is the
subsequent average over many disorder realizations in order to get a
reasonably smooth line shape. This can be an expensive task limited
by both computing time and memory resources. Therefore, it is
compelling to look for a scheme such as CPA which avoids large scale
computations but the CPA needs to be adapted to spatial correlation
as done in the present paper.

For excitons in a mixed semiconductor crystal, the basic random
quantity $U(\r)$ is the local energy gap due to $A$ and $B$ atoms
placed uncorrelated on the lattice sites (see \Eq{Smoothing}). In
this case, the smoothing function $W(\r)$ is determined by the
exciton relative wave function. Assuming for the 1s ground-state
exciton the standard exponential expression, we have
\be \label{Wexp} W(\r) \propto \sigma e^{-r/\xi} \ee
where the potential correlation length $\xi$ is proportional to the
exciton Bohr radius $a_B$. In semiconductor nanostructures as
quantum wells (quantum wires), an additional source of disorder is
the fluctuation of the well width (the wire cross section), which
leads to a variation of the confinement energies. If the length
scale of these fluctuations is smaller than the exciton Bohr radius,
it can be approximated again as a delta-correlated random process
$U(\r)$. The smoothing function $W(\r)$ is then the appropriate
relative exciton wave function in lower dimensions.

In the case of electrons moving in the potential of densely spaced
randomly distributed impurities, $W(\r)$ is proportional to the
(screened) potential of a single impurity. In Appendix A, we give
explicit expressions for exponential and for Gaussian type of the
potential correlation, and are listing more details taken from
Ref.~\cite{TakaRev}.

A more sophisticated disorder-related effect is quantum-mechanical
level repulsion of excitons which can be seen in time-resolved
Rayleigh scattering \cite{SavonaZim} or spectral correlation
spectroscopy \cite{Lienau}. While these effects are of two-particle
nature we concentrate in the present paper on the one-particle level
which is sufficient to describe the absorption line shape of
excitons. For a review on our work on excitons and disorder in
nanostructures see Refs.~\cite{TakaRev,RungeSSP}. Quite recently, we
were able to do simulations for optical spectra without the
separation into relative and cm exciton motion \cite{Grochol}.

\section{Formulation of the problem} \label{Formulation}

Let us consider a single particle (electron, exciton,...) with mass
$M$ which moves in a random potential $V(\r)$. The one-particle
Green's function obeys the inhomogeneous Schr\"odinger equation
\be \label{Schroe} \left( -\frac{\hbar^2}{2M} \DeltaR + V(\r) -
\hz \right) \G(\r, \r', z)
  = - \delta(\r-\r')\, ,\ee
with the complex energy $\hz$. This problem is usually called {\em
Continuous Anderson model}. The one-particle density of states is
defined as
\be \label{DOS} \rho(\omega) = \frac{1}{\Omega}
  \left\langle \int  d\r \, \Im \G(\r, \r, \omega - i0) \right\rangle \, .
\ee
Further, we are interested in the spectral function $A_\k(\omega)$
\be \label{Spec} A_\k(\omega) = \frac{1}{\Omega} \left\langle \int
d\r \, d\r' \, e^{i\k(\r-\r')} \Im \G(\r, \r', \omega - i0)
  \right\rangle \, .
\ee
($\Omega$ is a normalization volume). For excitons, $\r$ has to be
understood as the center-of-mass coordinate, and the appropriate
mass is $M = m_e + m_h$. The line shape of the excitonic absorption
(into the 1s ground state) is given by the spectral function at $\k
= \k_{light} \approx 0$.

The angular brackets in Eqs.\,(\ref{DOS} and \ref{Spec}) represent
the average over all realizations of the random potential $V(\r)$.
We take the average potential as zero of energy (that is the band
edge of the virtual crystal). Therefore, the disorder potential has
zero mean value. Further, we assume it to be Gauss distributed with
variance $\sigma$, and define its spatial correlation via
\be \langle V(\r)\rangle = 0 \, ,\quad \langle V(\r) \, V(\r')
\rangle = \sigma^2 \, g(\r-\r') \, . \ee
The correlation function $g(\r)$ is therefore normalized to
$g(\r=0) = 1$. All higher order correlations are assumed to
factorize. Such type of potentials can be generated by
integrating with a smoothing function $W(\r)$ over a
delta-correlated basic random process $U(\r)$
\bea \label{Smoothing}V(\r) & = & \int d\r' \, W(\r - \r') \, U(\r') \, , \\
\langle U(\r) \, U(\r') \rangle & = & \delta(\r-\r') \, . \nonumber
\eea
The correlation function is related to $W$ via the convolution
\be \sigma^2 g(\r) = \int d\r' \, W(\r + \r') \, W(\r') \, . \ee
Let $W(\r)$ decay on the scale $\xi$. Then, the resulting random
potential varies smoothly over distances of order $\xi$, and the
correlation of the potential can be quantified by the kinetic energy
on the length scale $\xi$
\be E_c = \frac{\hbar^2 \xi^{-2}}{2M} \, .\ee
Obviously, there are only two energies in the theory: The
correlation energy $E_c$, and the disorder strength $\sigma$. One of
them can be taken as energy scale. Given a certain type of
correlation (exponential or Gaussian), the spectral function (or
exciton line shape) depends therefore only on the single parameter
$E_c/\sigma$.

\section{Correlated CPA} \label{CorrelatedCPA}

The basic idea of the CPA is sketched as follows \cite{Soven,VKE}:
The entire space is divided into a central cluster where the
disorder average has to be performed exactly, and into the remaining
outer part which is described as an effective medium with averaged
(but complex) potential, respectively, self-energy $\Sigma(z)$. For
the determination of $\Sigma(z)$, the Green's function of the
effective medium $G(z)$ is put equal to the cluster average of the
full Green's function (cluster embedded into the effective medium).
In the conventional CPA, the central cluster is reduced to a single
site, and the self-consistency condition for $\Sigma(z)$ can be
formulated as a vanishing of the averaged $t$-matrix (Soven's
equation, \cite{Soven}). In a correlated potential, the abrupt
change between the on-site potential $V(\r=0)$ and the self-energy
$\Sigma(z)$ outside is in contrast to the smooth behavior of any
potential realization. Choosing a larger cluster is expected to
improve, but -- apart from analyticity problems encountered -- the
abrupt change at the cluster boundary remains unaffected. In order
to illustrate this point we look for the conditional probability to
find a potential value $V_1$ at a distance $\r$ keeping $V_0 \equiv
V(\r=0)$ fixed:
\be P(V_1, \r | V_0) = \frac{1}{\sqrt{2 \pi} \sigma_1(\r)}
  \exp \left(-\frac{(V_1 - V_0g(\r))^2}{2 \sigma_1^2(\r)} \right) \, .
\ee
The most probable value of $V_1 = V(\r)$ is therefore given by
\be
 \left. V(\r) \right|_{prob.} = V_0 \, g(\r)\, ,
\ee
and distributed with a variance $\sigma_1^2(\r) = \sigma^2 (1 -
g^2(\r))$ increasing with distance towards $\sigma$ (see \Fig{1} for
a visualization). The main idea of the present CPA extension is to
use a {\em smooth} interpolation between the central "most probable"
potential shape and the effective medium outside which is assumed to
have a nonlocal self-energy,
\be \label{SmoothPot} V(\r) \Rightarrow V_0 \, g(\r) \,
\delta(\r-\r') +
  \Sigma(\r-\r',z) (1 - g(\r')) \, .
\ee
In this way, any sharp break is avoided, and the effective
Schr\"odinger equation reads
\bea \left( -\frac{\hbar^2}{2m} \DeltaR + V_0 g(\r) - \hz \right)
\G(\r, \r'', z) & + &\\
 \int d\r' \,\Sigma(\r-\r',z)\left(1-g(\r')\right)\G(\r', \r'', z)& = & -
\delta(\r-\r'') \, .\nonumber \eea
\begin{figure}
\includegraphics*[angle=-90,width=70mm]{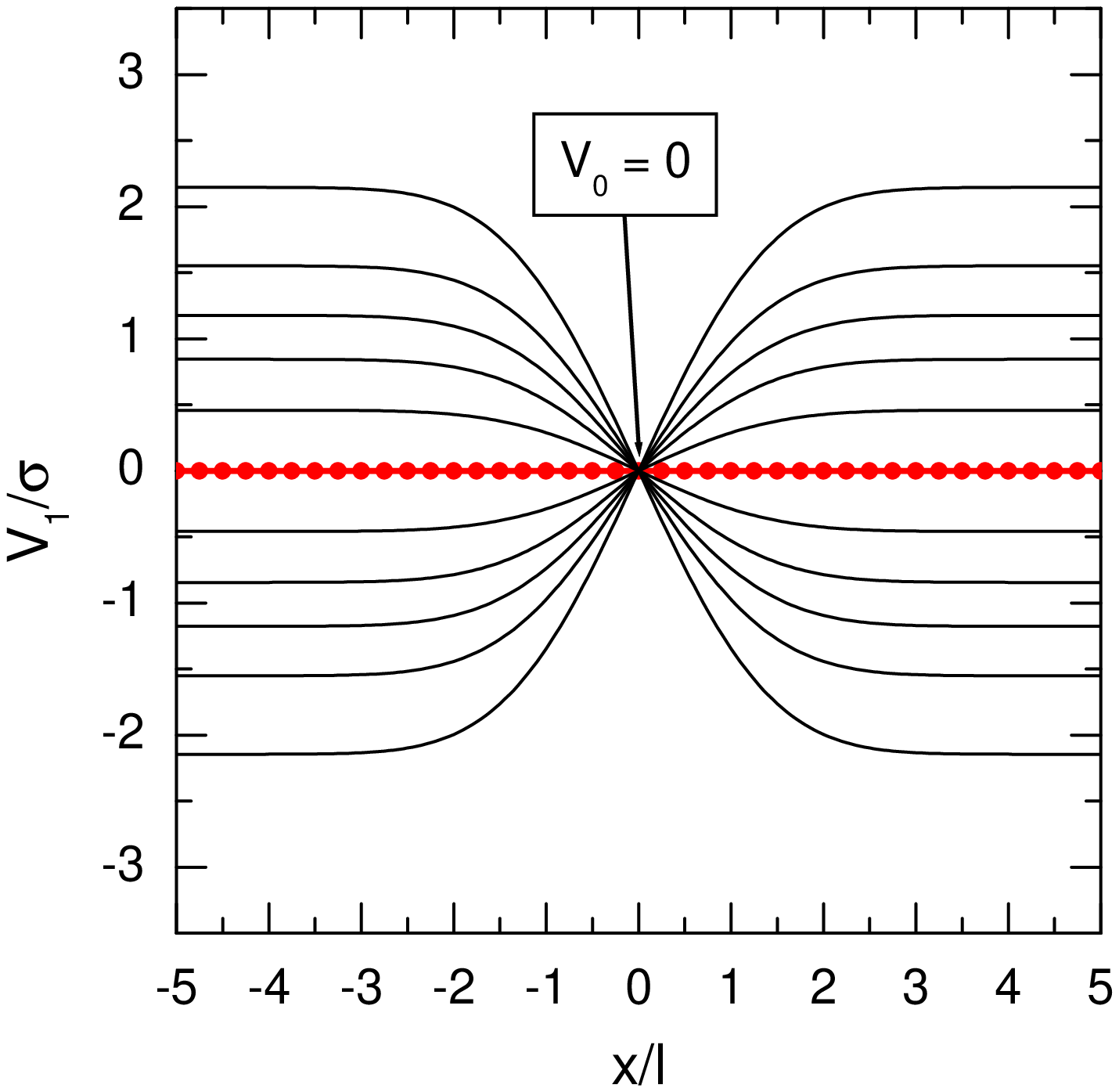}
\includegraphics*[angle=-90,width=70mm]{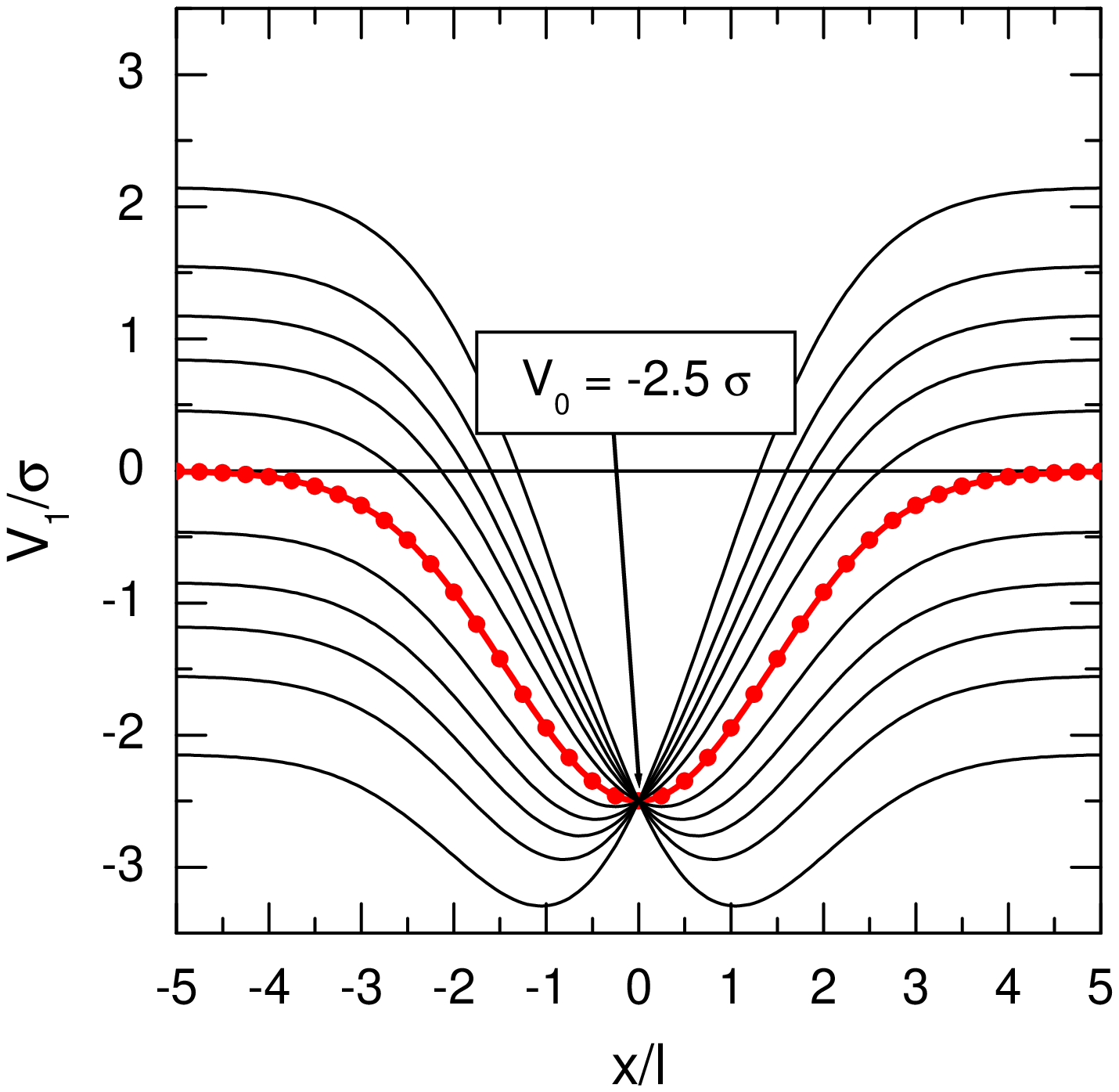}
\caption{\label{Fig1}(color online) Probability distribution of the
Gauss correlated potential around a fixed potential value at $x=0$
(top: $V_0 = 0$, bottom: $V_0 = - 2.5\,\sigma$). The curve with dots
marks the most probable value $V_0\,g(x)$. Other lines refer to
probabilities 0.9, 0.7, 0.5, 0.3, 0.1 relative to the maximum (at
each $x$ value).}
\end{figure}
Fourier transformation into reciprocal space yields
\bea  \label{Fourier} (\ek + \Sk - \hz) \, \G_{\k \k'}(z) &+ & \\
  +\,(V_0 - \Sk)\sum_{\kt} g_{\k \kt} \, \G_{\kt\k'}(z) &=& - \delta_{\k\k'} \,
  ,\nonumber
\eea
with dispersion $\ek = \hbar^2k^2/2M$. The Fourier transform of the
potential correlation function, $g_\k$, can be expressed via the
smoothing function as
\be \label{gq} \sigma^2 g_\k = \left| W_\k \right|^2 \; , \quad W_\k
= \frac{1}{\sqrt{\Omega}} \int d\r \, e^{-i\k\r} \, W(\r) \, .\ee
The normalization $g(0) = 1$ transforms into $\sum_\k \gk = 1$.
\Eq{Fourier} can be written as
\be \label{Fullq}
 \G_{\k \k'}(z) = G_\k(z) \left[ \delta_{\k\k'} +
   (V_0 - \Sk)\sum_{\kt} g_{\k \kt}\, \G_{\kt \k'}(z) \right]
\ee
introducing the Green's function of the effective medium
\be \label{GF}
 G_{\k}(z) = \frac{1}{\hz - \ek - \Sk}
\ee
which is diagonal in momentum space. In contrast, the Green's
function \Eq{Fullq} is nondiagonal since we have located the central
potential $V_0$ at $\r = 0$ which has spoiled the translational
invariance. Placing $V_0$ at position $\r_0$ leads simply to the
Green's function $\G(\r-\r_0, \r'-\r_0, z; V_0)$ (for clarity, we
have marked the dependence on $V_0$). Therefore, the embedding into
the effective medium should contain two steps: (i) an integration
over $\r_0$ and (ii) an average over the (Gaussian) distribution of
the central potential
\be \label{GaussV0} P(V_0) = \frac{1}{\sqrt{2 \pi} \sigma} e^
{-V_0^2/2\sigma^2} \, . \ee
The self-consistency condition in the spirit of the CPA reads
therefore
\be G(\r-\r', z) = \frac{1}{\Omega} \int\! d\r_0 \int\! dV_0 P(V_0)
\G(\r-\r_0, \r'-\r_0, z; V_0)\ee
or after Fourier transformation
\be \label{Soven} \left\langle \G_{\k \k}(z; V_0)
\right\rangle_{V_0} = G_{\k}(z) \, . \ee
Note that only the diagonal elements of $\G_{\k \k'}(z)$ enter the
CPA condition.

In order to avoid the solution of the matrix equation \Eq{Fullq} we
specify our Ansatz further by factorizing the momentum dependence as
\be  \label{Factorize} \G_{\k \k'}(z; V_0) \Rightarrow G_\k(z)\,
H_{\k'}(z; V_0) \, .\ee
Then, \Eq{Fullq} at $\k = \k'$ can be solved immediately, and the
self-consistency condition \Eq{Soven} converts into  $\langle
H_{\k}(z; V_0) \rangle_{V_0} = 1$. Explicitly, we have
\be \label{CPA} \left\langle \frac{1}{1 - (V_0 - \Sk)R_\k(\hbar
z;\{\Sigma\})} \right\rangle_{V_0} = 1 \ee
with the auxiliary function
\be \label{Aux1} R_\k(\hbar z; \{\Sigma\})  = \sum_{\k'} \gkk
G_{\k'}(z) = \sum_{\k'} \frac{\gkk}{\hz-\ekl -\Skl} \, . \ee
The argument for chosing the present Ansatz
Eqs.(\ref{SmoothPot},\ref{Factorize}) was to exactly reproduce three
well-known limiting cases:

(i) White noise: For an uncorrelated potential ($\gk = const.$), the
auxiliary function \Eq{Aux1} and therefore the self-energy do not
depend on momentum. Thus, the conventional CPA is fully recovered
which reads \cite{VKE}
\be  \label{StandardCPA} \left\langle \frac{1}{1 - (V_0 - \Sz)
\frac{1}{\cal N}
\sum_{\k}
\frac{1}{\hz-\ek - \Sz}}
\right\rangle_{V_0} = 1 \, . \ee
Here, a finite band width has to be implemented by restricting the
sum over $\k$ to the first Brillouin zone, having ${\cal N}$
discrete momentum points. The parabolic dispersion $\ek$ must be
replaced by a tight-binding expression, and one arrives at the CPA
of the \textit{discrete Anderson model} with site disorder.

(ii) Classical limit: If the correlation has infinite range, we have
$g(\r) = 1$ ($\gkk = \delta_{\k\k'}$), and the solution of \Eq{CPA}
is simply
\be \left\langle \frac{\hz - \ek - \Sk}{\hz - \ek - V_0}
\right\rangle_{V_0} = 1 \ee
with the result
\be G_{\k}(z) = \left\langle \frac{1}{\hz - \ek - V_0}
\right\rangle_{V_0} \, . \ee
The Green's function mirrors the potential distribution displaced
by $\ek$ as expected in this case.

(iii) Second Born approximation: For weak coupling, \Eq{CPA} can be
expanded into a geometrical series
\be \label{BornSeries} \sum_{j=0}^{\infty} \left\langle
\left(V_0-\Sigma_\k(z)\right)^j \right\rangle_{V_0} R^j_\k(\hbar z;
\{\Sigma\}) = 1 \, .\ee
Up to second order ($j\le 2$), we need the disorder averages
$\langle V_0 \rangle = 0$ and $\langle V_0^2 \rangle = \sigma^2$.
This results in leading order to the self-energy
\be \label{Self2B} \Sigma^{2B}_{\k}(z) = \sigma^2 R_\k(\hbar z;
\{\Sigma^{2B}\}) \equiv \sum_{\k'} \frac{\sigma^2\,\gkk}
    {\hz-\ekl - \Sigma^{2B}_{\k'}(z)}
\ee
which is just the self-consistent second Born approximation, also
called "random coupling method" \cite{Kraich}. This agreement was
the guiding principle for selecting the appropriate "switch on" of
the self-energy in \Eq{SmoothPot}. We were starting with $V_0 g(\r)
\delta(\r-\r') + \Sigma(\r-\r', z) (1 - \hat{g}(\r'))$ with a yet
unspecified function $\hat{g}(\r')$. Consequently, a second
auxiliary function $\hat{R}$ appears as factor to the (explicit)
self-energy in \Eq{CPA} and \Eq{BornSeries}, while $V_0$ stays with
the original $R$ ($\hat{R}$ is defined like \Eq{Aux1} but contains
$\hat{g}_\k$ instead of $g_\k$). In second Born quality, the
self-energy reads now $\Sigma^{2B} = \sigma^2 R^2/\hat{R}$ (with
arguments suppressed). The comparison with \Eq{Self2B} dictates
$\hat{R} \equiv R$ that is $\hat{g} \equiv g$, which justifies {\em
a posteriori} the choice made in \Eq{SmoothPot}.

An application of the self-consistent second Born approximation to
excitons in disordered semiconductors can be found in
Ref.~\cite{Stroucken}. In diagram language, this self-energy
consists of the simplest diagram where the averaged Green's function
appears just once. Although being an approximation with inferior
quality (see Sec.~\ref{Results}), already here a self-energy comes
out which depends continuously on momentum (the frequency $z$ plays
the role of a parameter). Other approaches like Jarrell's DCA
\cite{Maier} -- while much more sophisticated with respect to local
correlations -- fail to reproduce the self-consistent second Born
approximation in the appropriate limit, coming up with a
discontinuous momentum dependence of the self-energy.

The coupling of self-energies at different $\k$ points in
Eqs.\,(\ref{CPA} and (\ref{Aux1}) complicates the numerics since the
self-consistency can be achieved only iteratively. Replacing $\Skl$
by $\Sk$ in \Eq{Aux1} reduces the problem to a simple root-searching
task in the complex plane for the single quantity $\Sk$ (at given
values of $\k$ and $z$)
\be \label{CPAdiag} \left\langle \frac{1}{1 - (V_0 - \Sk)R_\k(\hbar
z-\Sk)} \right\rangle_{V_0} = 1 \ee
with the auxiliary function now written as
\be \label{Aux2} R_\k(\hbar z)  = \sum_{\k'} \frac{\gkk}{\hz-\ekl}
\, . \ee
We call this diagonal coherent-potential approximation and show in
Sec.\,V that the agreement with the simulation results is not much
sacrificed. With the self-consistently determined self-energy, the
spectral function is obtained as
\be \label{SpecDef} A_\k(\omega) = \left. \Im \frac{1}{\hz - \ek -
\Sk} \, \right|_{z=\omega - i0}  \ee
and the density of states can be calculated via
\be \rho(\omega) = \frac{1}{\Omega} \sum_\k A_\k(\omega) \, .\ee

\section{Analysis of moments} \label{Moments}

Before presenting numerical results, it is useful to provide a
somewhat deeper insight into the quality of the CPA method
presented here. We focus on the moments of the spectral function
defined as
\be \label{MomDef} M_n(\k) = \int \frac{d\hom}{\pi} A_\k(\omega)
\left(\hom - \ek\right)^n \, .\ee
As a benchmark, we are going to check the moments in CPA against the
exact ones. The first moments describe some basic features of the
spectral function: $M_0 = 1$ gives the normalization, $M_1$ is the
average position with respect to the bare dispersion $\ek$, and
$M_2$ is the overall width. More subtle is the effect of the next
moments, where $M_3$ can be related to the asymmetry (or skewness),
and $M_4$ to the importance of tails. A distinct advantage is that
the moments can be evaluated exactly -- as long as they are finite
(see Appendix B). The clue is the spectral representation of the
(disorder-averaged) Green's function
\be G_\k(z) = \int \frac{d\hom}{\pi} \frac{A_\k(\omega)} {\hz -
\hom} \ee
which leads to the asymptotic expansion
\be \label{Gmoments} G_\k(z) \simeq \sum_{n=0}^\infty
\frac{M_n(\k)}{\left(\hz - \ek\right)^{n+1}} \ee
for $|z| \rightarrow \infty$. The exact operator expression
\be G_\k(z) = \langle \k | \left(\hz - {\cal H}
\right)^{-1}|\k\rangle \ee
can be expanded in a similar way, and the moments are found as
disorder averages of powers of the Hamiltonian in \Eq{Schroe}
\be \label{MomH} M_n(\k) = \langle \k | \left({\cal
H}-\ek\right)^{n}|\k\rangle \, .\ee
For the evaluation, it is useful to work in reciprocal space with
\be \langle \k|{\cal H}|\k'\rangle = \ek \delta_{\k \k'}\, +\,
V_{\k-\k'} \, , \ee
where $V_\q$ is the potential of a single realization in Fourier
space. The first steps give immediately
\be M_0 = 1\; , \quad M_1 = \langle V_{\k - \k} \rangle = 0 \,
,\ee
since averages over odd powers of the potential vanish. In the next
step
\be M_2 = \sum_{\k'} \langle V_{\k - \k'}\,V_{\k' - \k}\rangle =
\sigma^2 \ee
holds, using $\langle V_{\q}\,V_{-\q'}\rangle = 
\sigma^2 \delta_{\q \q'} g_\q$ and the normalization of the
potential correlation. Interestingly enough, the first three
moments are universal, i.e. they do not depend on the correlation
energy, and have no $\k$ dependence either. In higher orders, a
mixing between kinetic energy and potential correlation appears,
which can be quantified by
\be K_n(\k) = \sum_{\k'} \gkk \left(\ekl - \ek \right)^n \, .\ee
For the third moment, we obtain
\be \label{M3} M_3 = \sigma^2 K_1 \ee
which is again independent on momentum due to the mirror symmetry
$g_\q = g_{-\q}$. In the next order
\be \label{M4} M_4(\k) = \sigma^2 K_2(\k)\, + \, 3 \sigma^4 \,
,\ee
where the second term stems from the average over four potentials,
$\langle V^4(\r) \rangle = 3 \sigma^4$. We proceed up to the fifth
moment
\be \label{M5} M_5(\k) = \sigma^2 K_3(\k) \, + \, 10 \sigma^4 K_1
\, .\ee
Explicit expressions for different correlation type and
dimensionality are collected in Appendix C.

At still larger $n$, things are getting involved rapidly. By
dimensional arguments, $K_n(\k=0)\propto E_c^n$ and the moments at
$\k = 0$ have the following structure
\be M_n(0) = \sum_{j=1}^{[n/2]} a_{n, j} \,\sigma^{2j} \,E_c^{n-2j}
\ee
which shows up already in the first moments listed above. Only the
leading $\sigma$ power has a simple structure, namely $a_{n, j=n/2}
= 1\cdot 3\cdots (n-3)\cdot (n-1)$ for even $n$. Taken alone, these
terms would reconstruct just the Gaussian potential distribution
(classical limit). Our earlier work using the derivation of moments
is summarized in Ref.~\,\cite{RungeSSP}. The exact coefficients
$a_{n,j}$ and thus the moments have been generated using symbolic
manipulations. However, the numerical load increases exponentially
and we were restricted to $n\le 20$. On the other hand, the
numerical results gave clear evidence that much more moments (of the
order of thousands) are needed for a proper description of the
spectral function. Consequently, earlier attempts to work with an
infinite (but restricted) subset were not successful
\cite{GlutschMom}. In particular at large $E_c/\sigma$, even
negative portions of the spectral function may appear.

To complete the comparison, we are now evaluating the moments in
CPA. First, the general relation between moments of the spectral
function and of the self-energy is established. In analogy to
\Eq{Gmoments}, the self-energy moments $S_l(\k)$ are defined by
\be \label{Smoments} \Sk \simeq \sum_{l=0}^\infty
\frac{S_l(\k)}{\left(\hz - \ek \right)^{l+1}} \, .\ee
We have no constant term in this expansion since the band edge of
the virtual crystal was taken as zero of energy.  Inserting
\Eq{Smoments} into \Eq{GF} gives together with \Eq{Gmoments}
\be \label{Recursion} M_0=1 \, , \quad M_1=0 \, , \quad M_n(\k) =
\sum_{l=0}^{n-2} S_l(\k)\, M_{n-l-2}(\k) \, .\ee
A similar recursion relation between moments holds for the electron
gas \cite{Vogt} -- with Coulomb interaction instead of disorder. The
auxiliary function \Eq{Aux2} decays at least as $1/E$ where $E
\equiv \hz - \ek$ in the following. Keeping terms up to $E^{-5}$, we
restrict the expansion \Eq{BornSeries} of the defining CPA equation
\Eq{CPAdiag} to $j \le 5$ and perform the disorder average over the
"central" potential $V_0$ using
\be \langle V_0^2 \rangle = \sigma^2 \, , \quad \langle V_0^4
\rangle = 3\sigma^4\ee
(odd orders do vanish). After division with $R$ we arrive at
\be \label{SigmaExp} \Sigma = \left( \sigma^2 + \Sigma^2 \right)R
- 3\sigma^2 \Sigma R^2 + 3 \sigma^4 R^3 \ee
which is accurate up to $E^{-4}$. Therefore, thanks to
\Eq{Recursion}, we can obtain the spectral moments up to $M_5$ as
desired. In a next step, we write the auxiliary function needed
in \Eq{CPAdiag} as
\be R_\k(\hz-\Sk) = \frac{1}{E} \sum_{\k'} \frac{\gkk}{1 - (\ekl
- \ek + \Sk)/E} \ee
and expand this up to $1/E^4$
\be \label{RuptoE4} R_\k = \frac{1}{E} + \frac{K_1}{E^2} +
\frac{\sigma^2 + K_2(\k)}{E^3}+ \frac{3\sigma^2 K_1 +
K_3(\k)}{E^4}  \, . \ee
Now we are ready to equate powers of $E^{-n}$ in \Eq{SigmaExp}
iteratively. In successive steps using \Eq{RuptoE4} we obtain
\bea \label{allS} S_0(\k) & = &\sigma^2 \nonumber \\
S_1(\k) & = & \sigma^2 K_1  \\
S_2(\k) & = & \sigma^2 K_2(\k)  + \, 2\sigma^4 \nonumber \\
S_3(\k) &=& \sigma^2 K_3(\k) + 6 \sigma^4 K_1 \, . \nonumber \eea
Note that the first two orders in \Eq{allS} have been already
exploited to write the two last terms in \Eq{RuptoE4} in compact
form. The recursion \Eq{Recursion} gives
\bea \label{M234} M_2 & = & S_0 = \sigma^2 \nonumber \\
M_3 &=& S_1 = \sigma^2 K_1 \\
M_4(\k) &=& S_2(\k) + S_0^2 = \sigma^2 K_2(\k) + 3\sigma^4\, .
\nonumber \eea
Therefore, the correlated CPA generates the exact moments up to
the fourth order. Differences show up in the fifth order. Here,
the diagonal CPA version produces
\be \label{M5CPA} M_5^{\mathrm{CPAd}}(\k) = \sigma^2 K_3(\k) + 8
\sigma^4 K_1 \ee
while the correct prefactor of the second term should be 10 (see
\Eq{M5}). Using the more complicated original form \Eq{CPA} gives
not a real improvement: Up to $M_4(\k)$ there is no change but the
mentioned numerical prefactor in \Eq{M5CPA} goes up to 9 only.

The self-consistent second Born approximation \Eq{Self2B} even fails
to give $M_4(\k)$ properly -- it has only $2\sigma^4$ instead of $3
\sigma^4$ in the last line of \Eq{M234}. Using diagram language, it
is exactly the crossing diagram in 4th order which is missing here
-- while it is contained in CPA.

\section{Results and discussion} \label{Results}

The quality of the present CPA method is judged using simulation
results which can be considered as exact solution of the continuous
Anderson problem. For the simulation, the $\r$ space is discretized
on a cubic mesh with step size $\Delta$, and for the $D$-dimensional
cube (side length $N\Delta$), periodic boundary conditions are
applied. In order to avoid discretization artifacts, the potential
should change smoothly along $\Delta$. We found $\xi/\Delta = 3$ as
a sufficient condition. The standard discretization of the Laplacian
operator in \Eq{Schroe} maps the problem to the discrete Anderson
model with correlated potential. The corresponding transfer energy
is given by $T = \hbar^2\Delta^{-2}/2M$. A straightforward
diagonalization would give all $N^D$ eigenvalues and eigenfunctions
but is restricted to unacceptable small sizes $N$ because of memory
size and computation time. Earlier, we had numerically solved the
corresponding time-dependent Schr\"odinger equation and generated
the spectral function by time Fourier transformation following
Glutsch \cite{GlutschTime}. Later on, we have implemented a
numerical generation of Chebyshev moments, too \cite{RungeSSP}.
Going further this way, we apply in the present work the powerful
Kernel polynomial method (KPM) as detailed in Ref.~\cite{Alvermann}.
In essence, successive moments of the Hamilton operator in terms of
Chebyshev polynomials are generated. The relation to the standard
moments \Eq{MomDef} is straightforward. However, the essential
difference is that here not the exact (disorder-averaged) moments
are generated but those of a given potential realization. The
Chebyshev coefficients are damped according to the Jackson algorithm
which gives in the spectrum a nearly Gauss-shaped line for each of
the eigenvalues. In this way, a smooth spectral function can be
obtained after adding up results of a sufficient number of
independent disorder realizations. We have carefully checked that
all these technical parameters are chosen such that no influence on
the final shape of the spectral function is seen. For a typical
calculation in $D=2$, a square grid with $N = 100$ has been used,
and an average over 3000 realizations was performed. In order to get
a reasonably smooth spectrum without too much broadening, 500--1500
Chebyshev moments were calculated for the Jackson algorithm. The
numerical effort was maximal in $D=3$, where 500 realizations for a
box with side length $N = 50$ have been added up.

\begin{figure}
\includegraphics*[width=60mm]{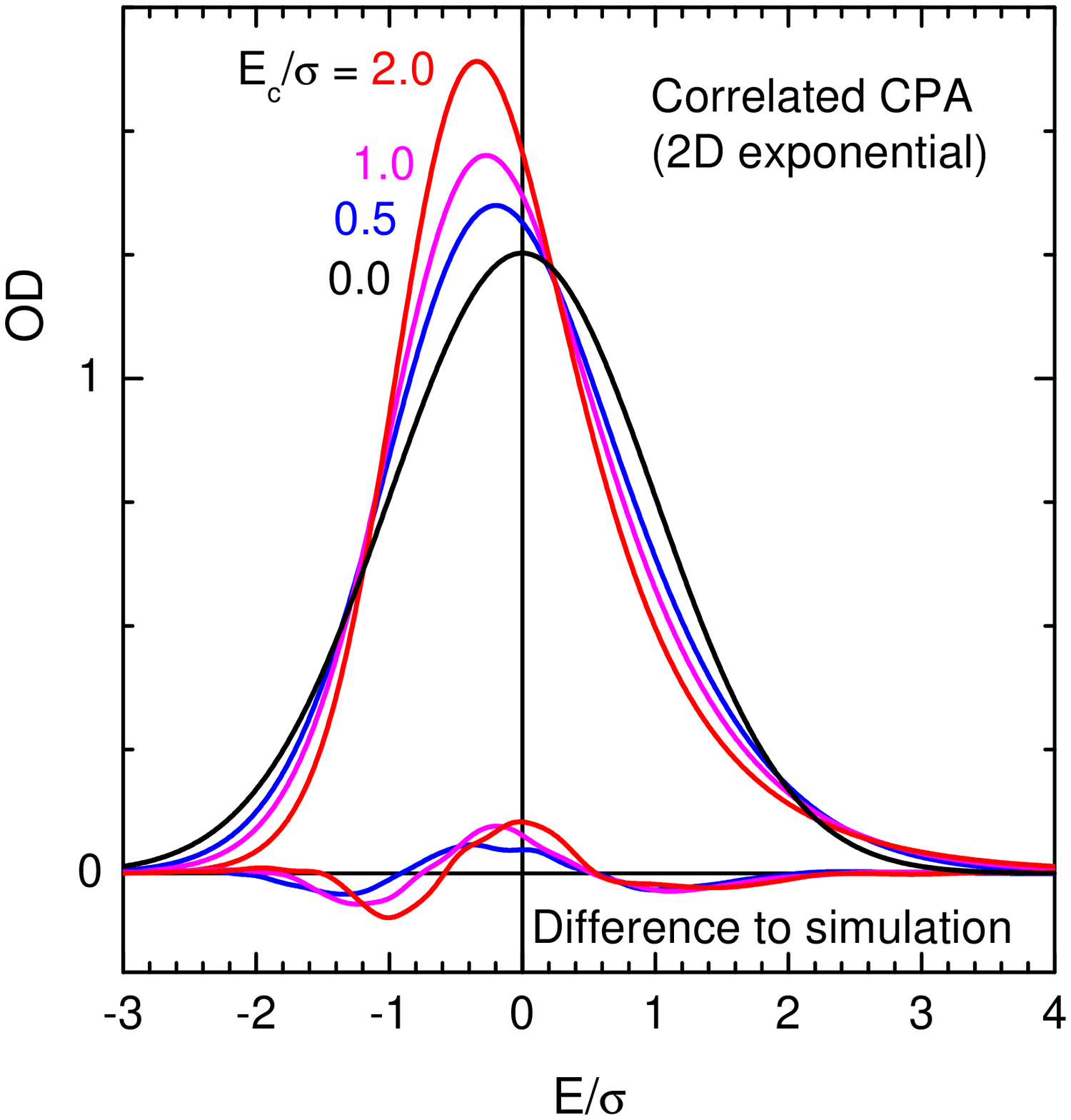}
\caption{\label{Fig2}(color online) Optical density for
exponentially correlated disorder in two dimensions. CPA
calculations in the diagonal version are shown for increasing ratios
$E_c/\sigma$ (increasing peak height). To judge the quality,
differences to the simulation results are shown on the same scale at
the bottom.}
\end{figure}
We begin with a comparison of results in $D=2$ which are relevant
for quantum wells with disorder. The correlation type is taken
exponential, and we concentrate here and in what follows on the
spectral function at $\k = 0$ which gives directly the inhomogeneous
broadening of the exciton absorption line, called {\em optical
density} in the following. Starting with the Gaussian potential
distribution in the classical limit ($E_c/\sigma = 0$), the curves
in \Fig{2} are getting more narrow and asymmetric for increasing
values of $E_c/\sigma$. This represents the motional narrowing on
the cm level. The correlated CPA deviates only slightly from the
numerically exact results, as visualized by the bottom curves.

\begin{figure}
\includegraphics*[angle=-90,width=60mm]{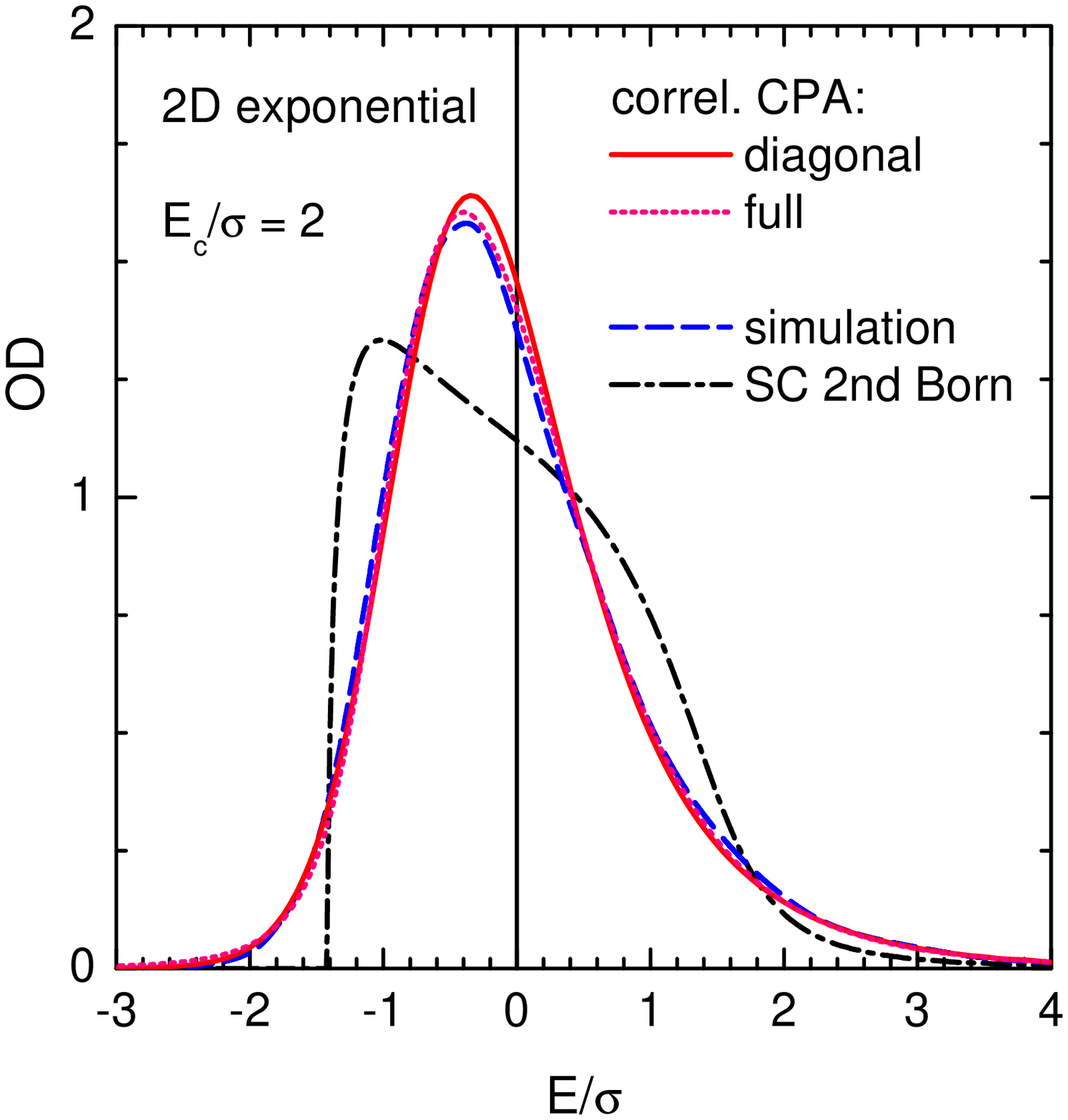}
\caption{\label{Fig3}(color online) Optical density for
exponentially correlated disorder in two dimensions at $E_c/\sigma =
2$. Different levels of approximation are compared with the
simulation results (see text).}
\end{figure}
In \Fig{3}, we compare the case $E_c/\sigma = 2$ for different
levels of approximations. The self-consistent second Born
approximation (dashed-dotted) deviates markedly. Only its
high-energy asymptotics is reasonable since it is dominated by
perturbation theory. In particular, it fails completely in the
low-energy tail where a sharp (square-root) cutoff is produced. This
is an inherent feature of any diagrammatic expansion which leads to
a geometrical series in terms of the interaction (disorder). The CPA
does here pretty well since the final summation over the local
potential fluctuations brings in a true random feature. The full
version (dotted line, \Eq{CPA}) gives only a slight improvement
compared to the diagonal approximation (full line, \Eq{CPAdiag}).
However, the numerics for the full problem is much more involved
since at a given energy, the complete momentum-dependent self-energy
has to be brought to convergency. We succeeded only by using
acceleration, respectively, slowing down in the recursive
determination. On the other hand, the diagonal version needs only a
single zero search in the complex plane. Therefore, we show in all
the other figures exclusively results from the diagonal version.
Then, for the optical density, only $\Sigma_{\k=0}(z)$ has to be
determined self-consistently.

\begin{figure}
\includegraphics*[angle=0,width=60mm]{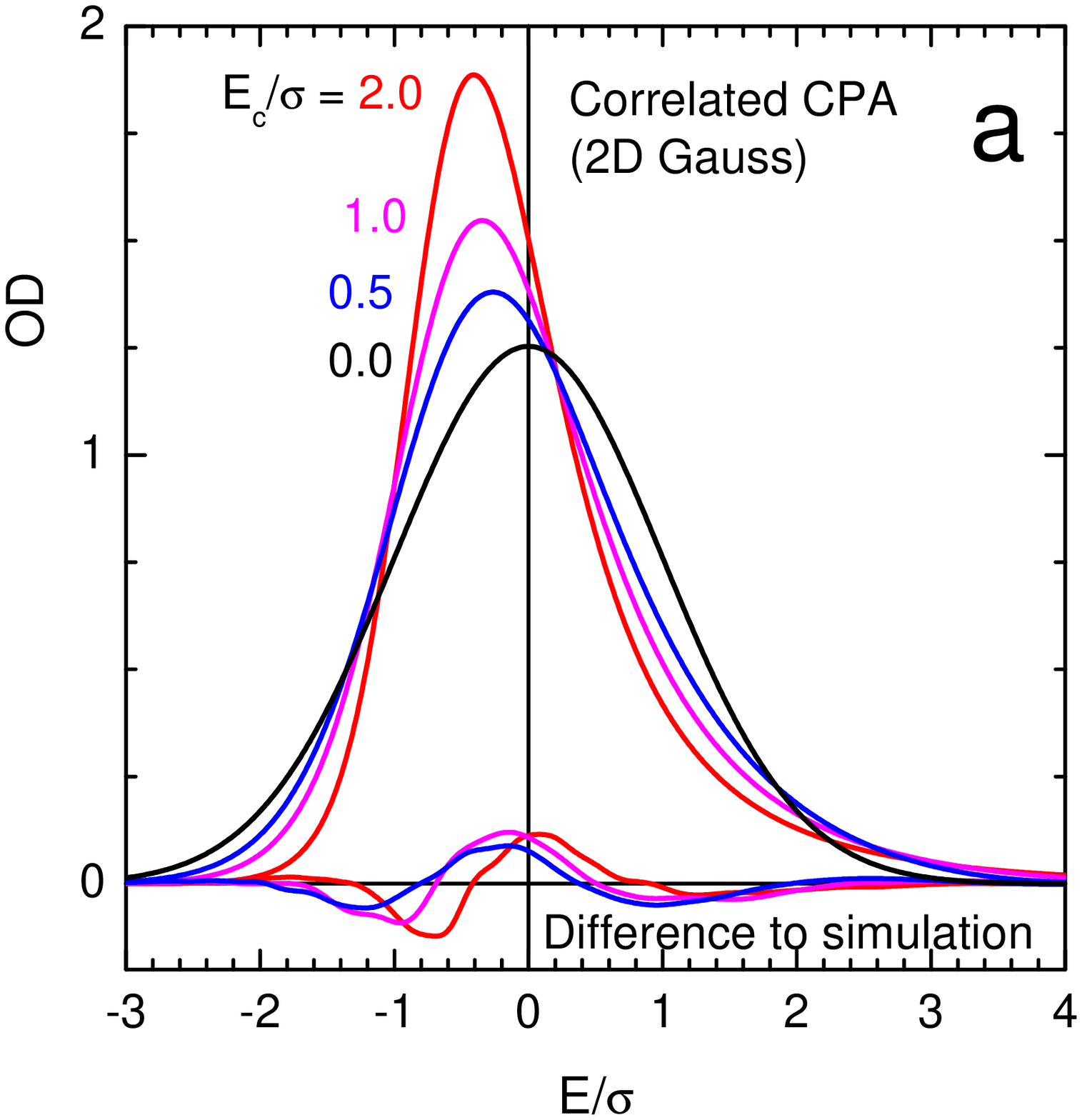}\\[2mm]

\hspace*{-5mm}\includegraphics*[angle=-90,width=65mm]{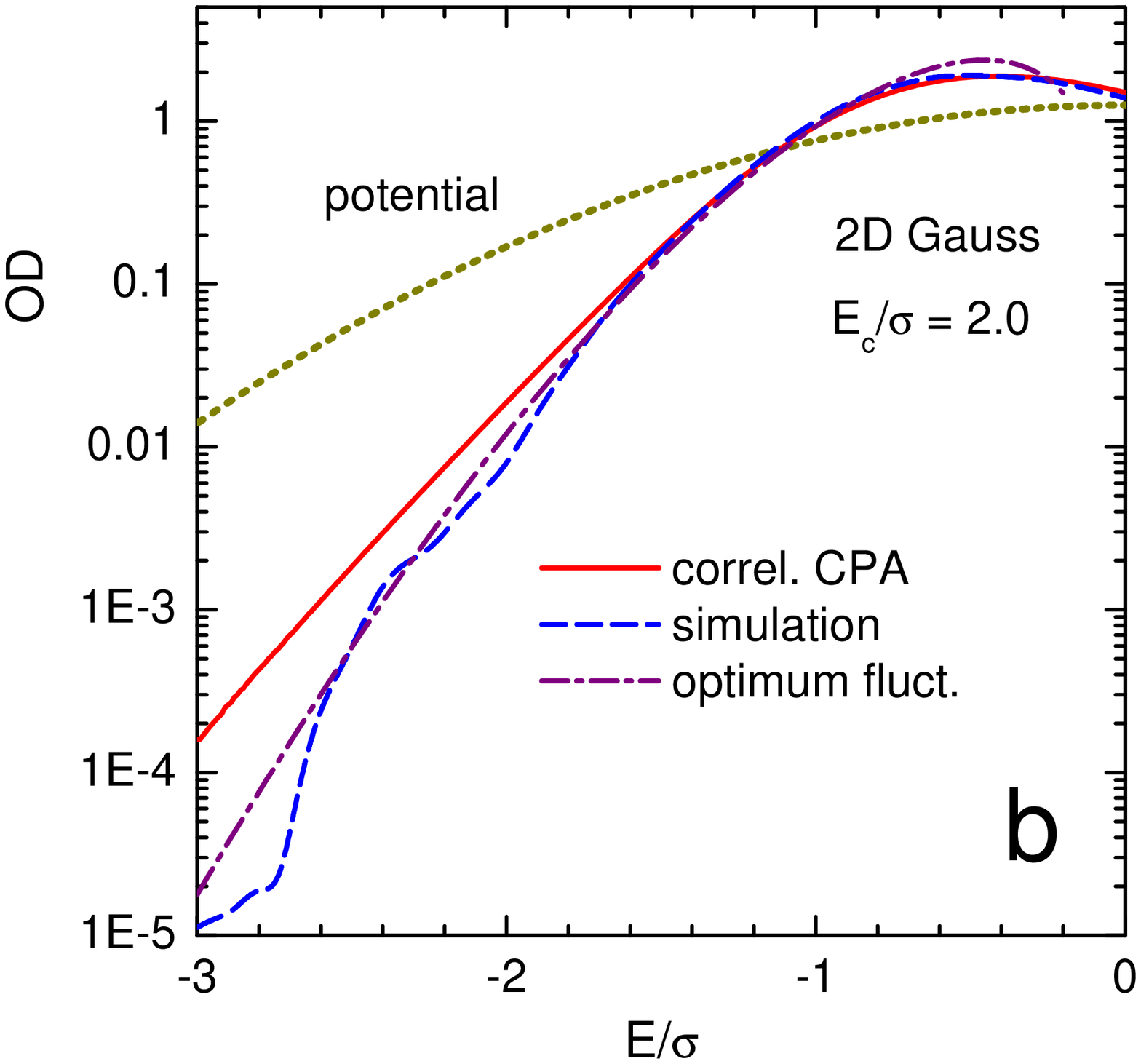}\\
\caption{\label{Fig4}(color online) Optical density for Gauss
correlated disorder in two dimensions. In addition to the linear
display in (a), the logarithmic display in (b) emphasizes the tail
region. The optimum fluctuation result (chained curve) has been
adjusted vertically.}
\end{figure}
It is pleasing to see how the quality of the different
approximations (self-consistent second Born, diagonal CPA, and full
CPA) goes in parallel to the number of moments which are exactly
reproduced. Taking all other parameters unchanged, a correlation of
Gauss shape leads to narrower spectral functions compared to the
exponential type (Fig.\,4(a)). The low-energy tail is shown on a
semi-logarithmic plot in Fig.\,4(b). The simulation is getting noisy
there since states deep in the tail are rare events but compares
well with the asymptotically strict result of the {\em optimum
fluctuation theory} \cite{OptFluc}, extended here to finite spatial
correlation \cite{RungeSSP}. While the correlated CPA follows
initially rather close, deep in the tail the spectral function is
somewhat overestimated.

\begin{figure}
\includegraphics*[angle=0,width=60mm]{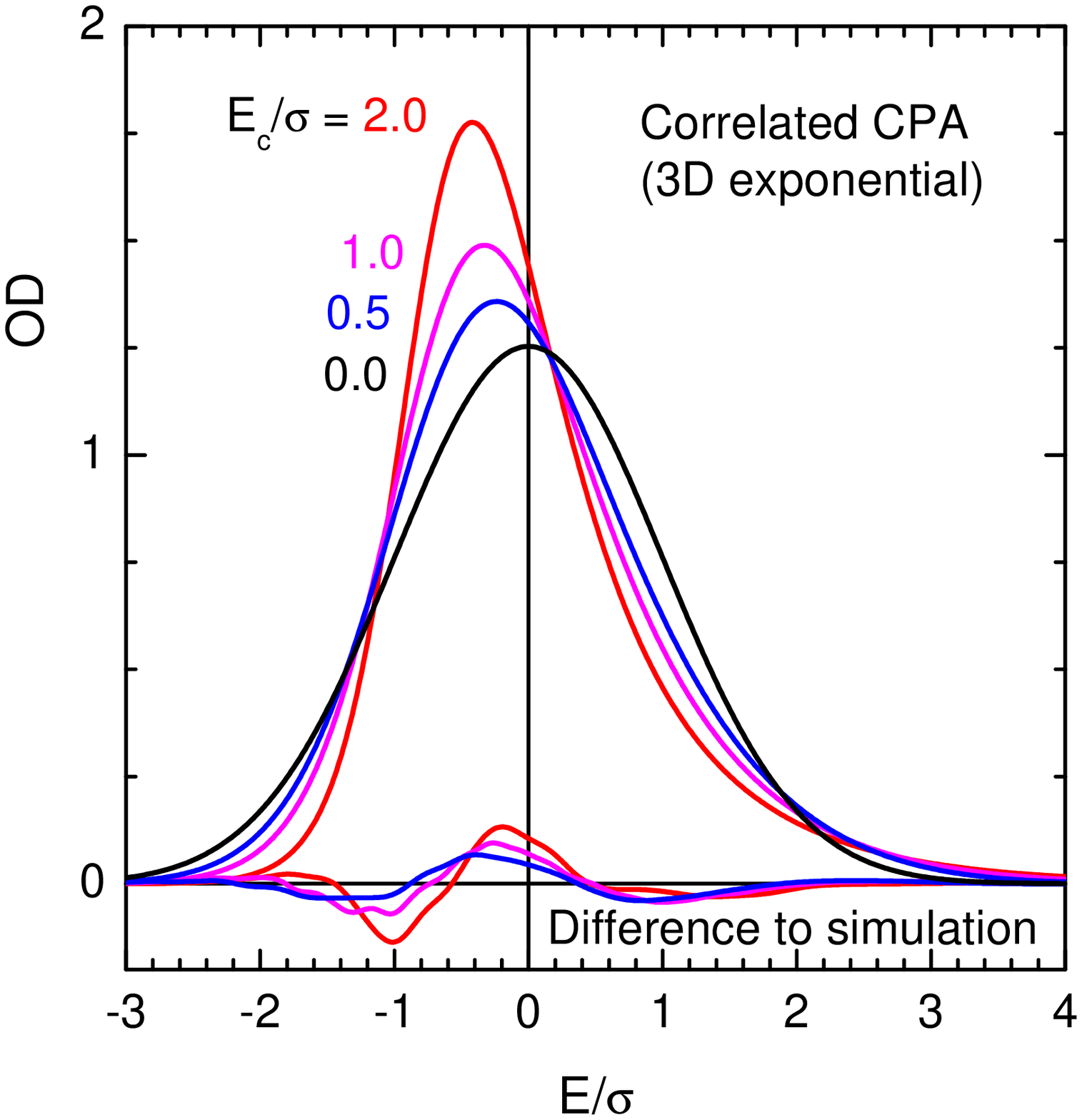}
\caption{\label{Fig5}(color online) Optical density for
exponentially correlated disorder in three dimensions.}
\end{figure}
When going from two to three dimensions (\Fig{5}), the line width at
a given value of $E_c/\sigma$ is reduced (cp. Fig.\,4(a)). This can
be related to the wave-function extension as -- loosely speaking --
an increased localization length. As well known, it is harder to
localize the wave function in $D = 3$ compared with $D = 2$,
assuming the same disorder strength and correlation type. Since we
are here interested in states around the lower band edge (these have
the dominant contribution to the $\k = 0$ spectral function), the
localization edge of the Anderson model in $D = 3$ is of no
relevance here.

\begin{figure}
\includegraphics*[angle=0,width=60mm]{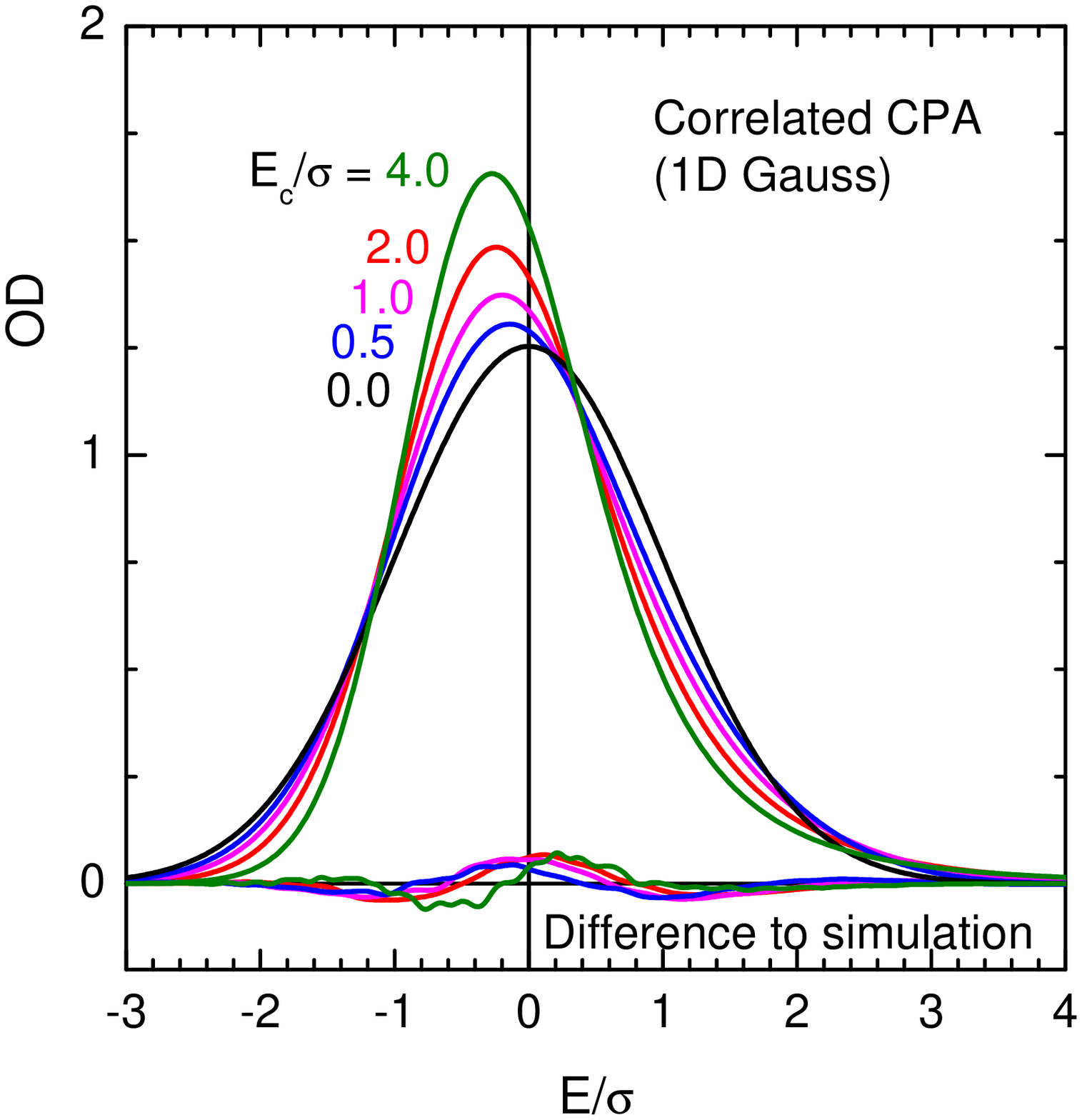}
\caption{\label{Fig6}(color online) Optical density for Gauss
correlated disorder in one dimension.}
\end{figure}
To complete the analysis, we show in \Fig{6} results for the
one-dimensional problem with Gauss correlated disorder. A final plot
(\Fig{7}) deals with the uncorrelated case in $D = 2$ which is the
realm of the standard single-site CPA in the discrete Anderson
model. The dispersion to be used here
\be \label{TB} \ek = 2T \sum_{j=1}^2 \left[1 - \cos(k_j \Delta)
\right] \ee
refers to a simple cubic lattice (${\cal N} = \Omega/\Delta^2$). The
deviations from the (numerically exact) simulation results are
definitely smaller than in the correlated cases (Figs.\,2 and 4).
This is in complete accordance with the moment analysis, since the
standard tight-binding CPA for uncorrelated disorder preserves the
exact moments even up to $M_7(\k)$ (we quote the first orders in
Appendix C, \Eq{TBmom}). Still, the CPA extension towards correlated
potentials is of reasonable quality, and provides therefore a new
tool for studying disorder problems in solid state physics.

The present method could be used for the discrete case with spatial
correlation as well. The potential generation \Eq{Smoothing} has to
be discretized as
\bea \label{SmoothDiscrete} V(\R) & = & \sum_{\R'}  W(\R - \R') \, U(\R') \, , \\
\langle U(\R) \, U(\R') \rangle & = & \delta_{\R \R'} \nonumber \eea
where $\R$ denote the lattice points. All Fourier transforms are
restricted to the first Brillouin zone, and in Eqs.(\ref{Aux1}) and
(\ref{Aux2}) $\ek$ has to be taken as tight-binding dispersion
again. However, an application to a {\em binary alloy} with spatial
correlation (clustering) of the chemical species is not possible: In
our method, the (local) potential must be Gauss distributed due to
the averaging process \Eq{SmoothDiscrete}, while the relevant
correlated potential for the alloy should be still a binary
quantity.
\begin{figure}
\includegraphics*[angle=0,width=60mm]{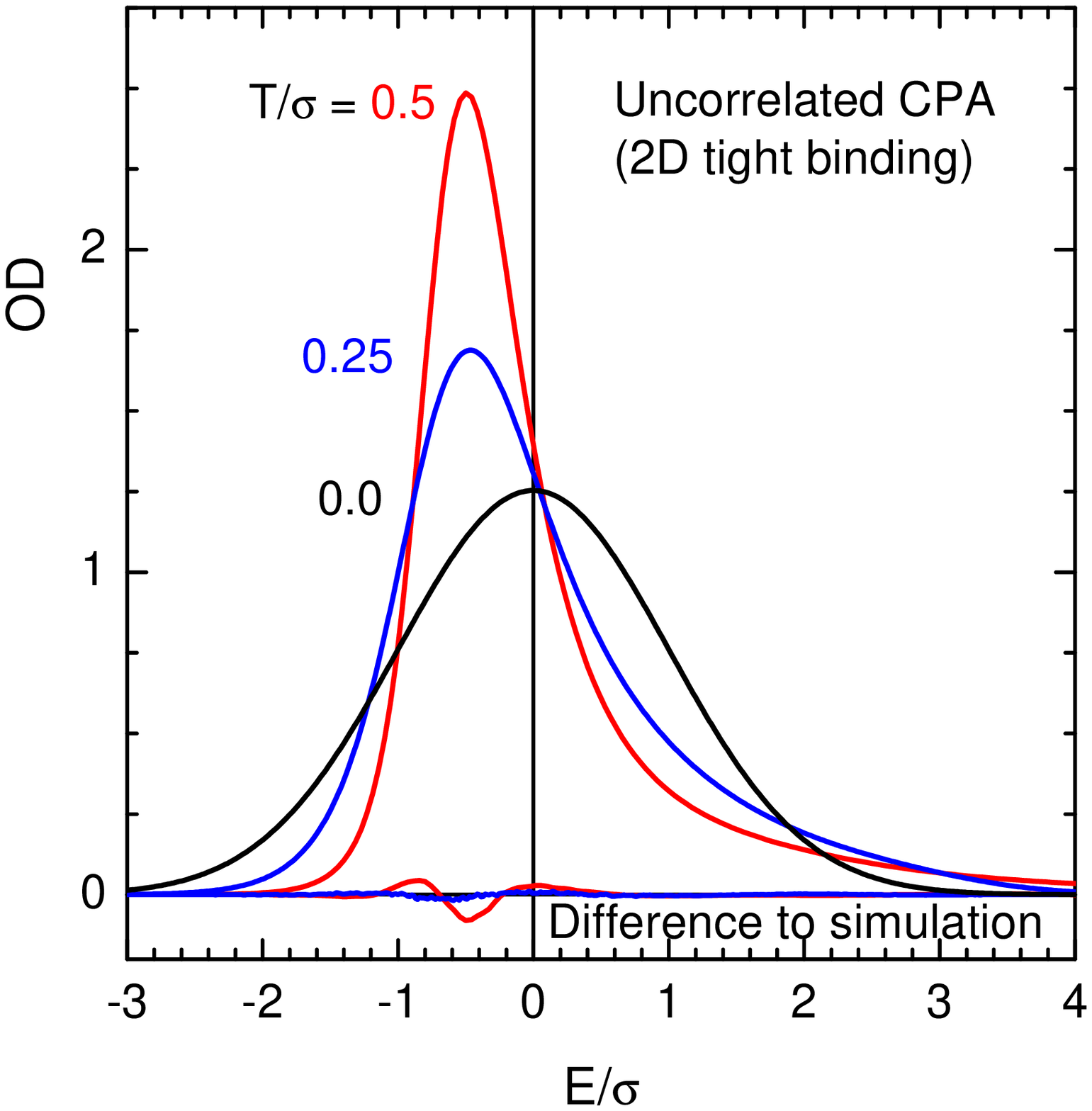}
\caption{\label{Fig7}(color online) Optical density (or zero
momentum spectral function) in a two-dimensional tight-binding
model. The standard CPA result for uncorrelated site disorder is
shown and compared with the simulation (here using 2048 Chebychev
moments).}
\end{figure}

\section*{Acknowledgments}

Enlightening discussions with Henry Ehrenreich in the early stage of
the work are gratefully acknowledged. We thank very much Erich Runge
for an ongoing exchange of ideas on disorder problems and for a
critical reading of the manuscript.

\appendix

\section{Analytic properties of the self-energy}

The proper Green's function of the averaged medium $G_\k(z)$ --
considered as a function of the complex-frequency argument $z$ --
should obey the following analytic properties (called Herglotz
properties, HP \cite{Mills}): It is analytic everywhere outside the
real axis, obeys the mirror symmetry $G_{\k}(z^*) = G^*_{\k}(z)$,
and its imaginary part is non negative in the lower half plane.
Using physics language, HP guarantee that the Green's function
refers to a causal system, and the spectral function (the jump on
the cut along the real axis, see \Eq{SpecDef}) is non negative, as
appropriate for the probability amplitude for adding a particle with
momentum $\k$ to the system. Due to the simple relation to the
self-energy \Eq{GF}, $\Sk$ should have HP as well. An alternative
formulation is (i) the validity of the spectral representation
\be \label{Herglotz} \Sk = \int_{-\infty}^\infty \frac{d
\omega}{\pi} \frac{s_\k(\omega)}{z - \omega}\ee
with (ii) a non negative spectral density $s_\k(\omega) \ge 0$.
Strictly speaking, a function with HP could have -- in addition to
the integral in \Eq{Herglotz} -- a constant and a term linear in
$z$. However, in the present case such terms are absent since the
self-energy has the virtual crystal as reference. Formally, they are
missing in the asymptotic expansion \Eq{Smoments} as well. In the
following we will show that the present extension of the CPA to
correlated disorder generates a self-energy with Herglotz
properties.

The mirror symmetry $\Sigma_\k^*(z) = \Sigma_\k(z^*)$ is obvious
since -- apart from $\Sk$ and $z$ -- only real functions ($\ek$,
$g_\k$, $P(V_0)$ enter the CPA equations. To be definite we place
now $z$ into the lower half plane ($\Im z < 0$) and search for a
self-energy with $\Im\Sk \ge 0$. Then, it follows at once that the
auxiliary function $R_\k(\hbar z, \{ \Sigma\}) \equiv R$ defined in
\Eq{Aux1} has a positive imaginary part, $R_2 > 0$ (note that
$g_{\k-\k'}$ is non-negative, \Eq{gq}). The self consistency
condition \Eq{CPA} is written explicitly as
\be \label{CPAexpl} \int dV_0 \frac{P(V_0)}{1/R + \Sigma - V_0} =
R\ee
where $\Sigma \equiv \Sk$. The integral looks like a standard
spectral representation, but the imaginary part of the denominator
$D \equiv 1/R + \Sigma - V_0$ which is $D_2 = \Sigma_2 - R_2/|R|^2$
may cross zero at some curve in the lower half plane. This would
give rise to a non-analytic self-energy there but can indeed never
occur: The imaginary part of \Eq{CPAexpl} reads
\be - D_2 \int dV_0 \frac{P(V_0)}{|D|^2} = R_2\ee
with a strictly positive value of the integral since $P(V_0)$ is a
positive probability distribution. Therefore, a solution can never
have $D_2 = 0$ since $R_2 > 0$ as shown above. This completes the
proof that the self-consistent self-energy has HP. The diagonal
simplification of \Eq{CPAdiag} does not change any step of the
proof. In this case, the even stronger assertion $D_2 \le \hbar z_2
< 0$ holds since the inequality $\Im (1/R_\k(\tilde{z}) -
\hbar\tilde{z}) < 0$ can be proven quite generally ($\hbar\tilde{z}
\equiv \hbar z - \Sigma_k(z)$).

In order to see the desired behavior in the numerics as well, we
have been searching in the full complex $z$ plane for
self-consistent solutions of the self-energy. Indeed, the
self-energy was found to vary smoothly as a function of $z$
(crossing non-analytic points/curves would show up as a
discontinuity). Still, there could be more than one solution.
Assuming at large $|z|$ a vanishing self-energy as initial guess in
the root search, we were at least starting with the proper solution,
and have continued with the previous solution as start for the next
$z$ position. A branching of this solution into two is not possible
-- this would signal a non-analytic point. However, there could be
an accidental degeneracy with a second solution. Although this never
occurred in our calculations, we could imagine how to select
numerically the proper continuation (demanding the absence of breaks
in slope). We were not able to give a general proof for the
uniqueness of the solution, as done by Mills and co-workers in
Ref.\cite{Mills} for the substitutional alloy with uncorrelated
disorder, and in Ref.\cite{Mills84} for a chain with randomly placed
delta scatterers. A more practical proof for $\Sigma_\k(z)$ having
HP is to check that the generated function obeys the spectral
representation \Eq{Herglotz}. To do so it is sufficient to run the
calculation for $z = \omega  - i0$ only, since $s_\k(\omega) \equiv
\Im \Sigma_\k(\omega - i0)$. The result in \Fig{8} shows a complete
agreement between the direct result and the spectral form
\be \label{SpecForm} \Re\Sigma_\k(\omega) = \int_{-\infty}^\infty
\frac{d \omega'}{\pi} s_\k(\omega'){\cal P} \frac{1}{\omega -
\omega'}\ee
where $\cal P$ stands for the principal value. For comparison, we
have added in the figure exact self-energy results which follow from
the complex Green's function generated with our simulation technique
using \Eq{GF}.
\begin{figure}
\includegraphics*[width=60mm]{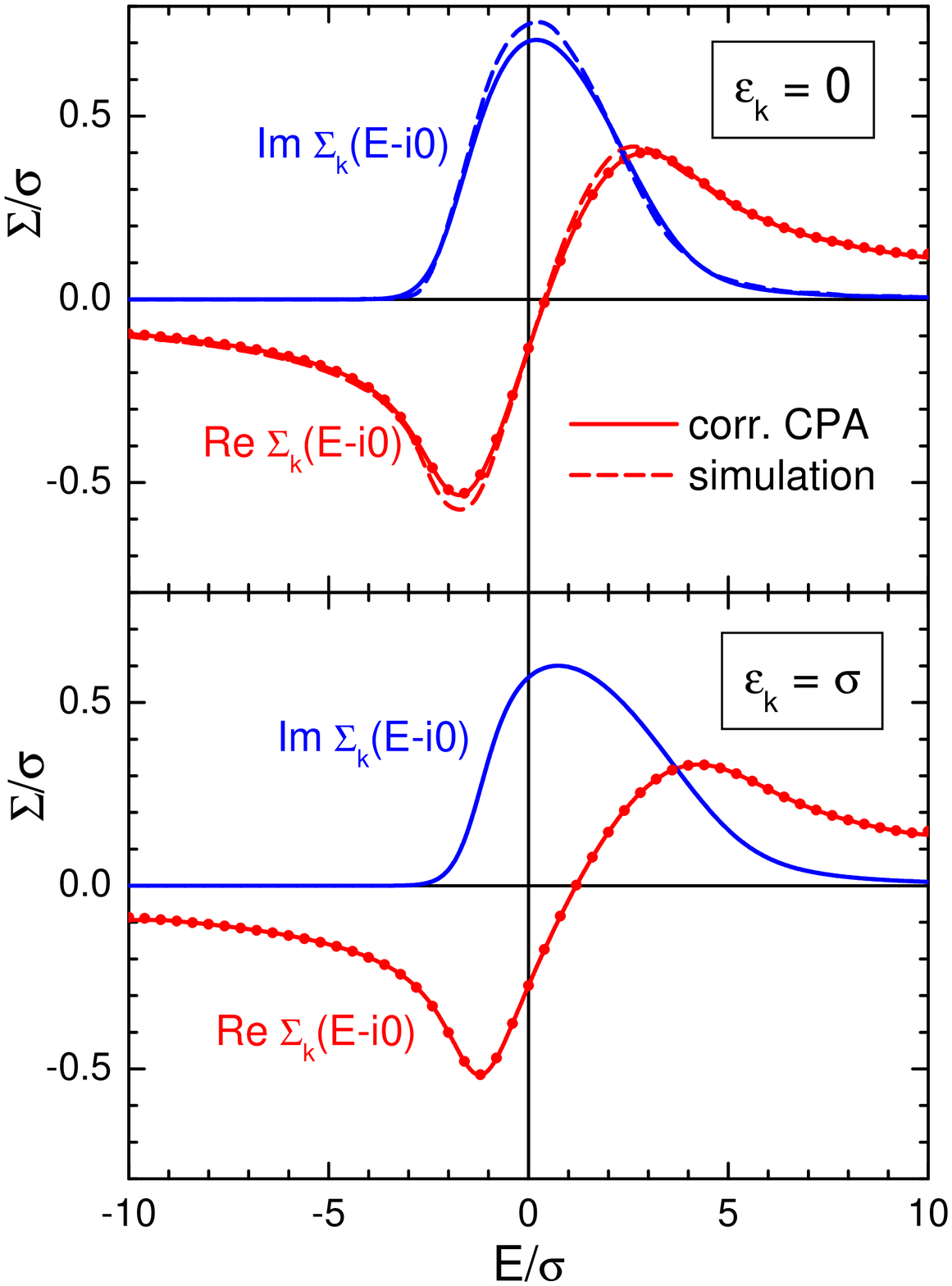}
\caption{\label{Fig8}(color online) CPA self-energy for exponential
correlation ($E_c/\sigma = 1$ in two dimensions) at zero momentum
(top) and at $\ek = \sigma$ (bottom). The dots give the real part
calculated via \Eq{SpecForm}. The exact self-energy for $\ek = 0$
from the simulation is shown dashed.}
\end{figure}

\section{Explicit expressions for different correlation types}

The probability distribution of the local potential value $V_0$ is
of Gauss type, \Eq{GaussV0}.  Therefore, the integral in
\Eq{CPAexpl} can be expressed using the complex error function
\be \label{ComplError} w(z) = \frac{i}{\pi}\int_{-\infty}^{+\infty}
\frac{e^{-t^2} dt}{z - t} \ee
resulting in
\be i\sqrt{\frac{2}{\pi}}\sigma R =
w\left(\frac{1}{\sqrt{2}\sigma}(1/R + \Sigma_\k(z)\right)\,.\ee
For the diagonal version, $R \equiv R_\k(\hbar z - \Sk)$ is
understood.
The specification of the potential correlation function depends on
the system under investigation.  If the potential is short-range
correlated, we approach the limit of a "white noise potential"
($g_\k = 1/\Omega$). This is the proper situation for the
traditional CPA where the self-energy is assumed to be site diagonal
(i.e. independent on momentum $\k$).

For the exciton case, it is important to note that electron and
hole are scanning the potential landscape on a different scale,
depending on their masses $m_e$ and $m_h$. For quantum wells,
local fluctuations of the well width $L_z$ lead to local shifts of
the band edges $E_a(L_z)$, and the smoothing function has the
following form \cite{TakaRev}
\be \label{Wr} W(\r) = h \zeta \sum_{a=e,h} \eta_a^2
\phi_{1s}^2(\eta_a \r) \frac{dE_a}{d L_z} \ee
with mass factors $\eta_e = M/m_h$ and $\eta_h = M/m_e$. The
parameters $h$ ($\zeta$) characterize typical height (lateral
extension) of the well width fluctuations (island size). The
exciton wave function can be taken hydrogen like, $\phi_{1s}(\r)
\propto \exp(-r/a_B)$, where $a_B$ is he appropriate Bohr radius
of the quantum well exciton. For equal electron and hole mass
($\eta_e = \eta_h = 2$), \Eq{Wr} reduces to a single exponential
dependence
\be W(\r) = \sqrt{\frac{2}{\pi}} \,\frac{\sigma}{\xi} \,e^{-r/\xi}
\ee
where $\xi = a_B/4$ and
\be \sigma = \sqrt{\frac{2}{\pi}} \frac{h \zeta}{a_B}(E'_e + E'_h)
\, .\ee
Performing the two-dimensional integrations we evaluate \Eq{gq}
with the result
\be g_\k = \frac{8\pi \xi^2}{\Omega} \frac{1}{(1+(k \xi)^2)^3} \,
. \ee
The auxiliary function $R_\k(\hz)$ \Eq{Aux2} at $\k=0$ reads
\be R_0(\hz) = \int_0^\infty\! \frac{dx}{\hbar z - E_c\, x}
\frac{2}{(1+x)^3} = \frac{s^2 + 4s + 3 + 2\log(-s)}{E_c (1+s)^3} \ee
where $s=\hbar z/E_c$ is the dimensionless complex energy.

The corresponding results for exponential correlation in quantum
wires (one dimension) are
\be g_\k = \frac{4\xi}{\Omega} \frac{1}{(1+(k \xi)^2)^2}\, ,\ee
\bea R_0(\hz) &=& \frac{4}{\pi} \int_0^\infty \frac{dx}
{(\hz - E_c\, x^2)(1+x^2)^2} \\
&=& \frac{s + 3 - 2/\sqrt{-s}}{E_c (1+s)^2} \, .\nonumber \eea
Finally, we apply \Eq{Wexp} to the bulk mixed crystal and get
\be g_\k = \frac{64 \pi \xi^3}{\Omega} \frac{1}{(1+(k \xi)^2)^4}
\ee
which is followed by
\bea R_0(\hz) &=& \frac{32}{\pi} \int_0^\infty \frac{x^2
dx}{\hbar z - E_c\, x^2} \frac{1}{(1+x^2)^4} \\
 &=& \frac{s^3 + 5s^2 + 15 s - 5 +16 \sqrt{-s}}{ E_c(1+s)^4}
\nonumber \, .\eea
For a Gauss-type correlation with characteristic length $\ell$, we
assume $W(\r) \propto \exp(-r^2/2\ell^{2})$ and obtain
\be  \label{GaussCorr} g_\k = \frac{1}{\Omega} (2 \sqrt{\pi}\ell)^D
e^{-k^2\ell^2} \, , \ee
where $D = 1, 2, 3$ is the spatial dimension. The correlation
energy is now defined as $E_c = \hbar^2/(2M\ell^2)$, and we have
to evaluate
\be R_0(\hz) = \frac{1}{E_c \pi^{D/2}} \int_{-\infty}^{+\infty}
d^D x \, \frac{e^{-x^2}}{s - x^2} \ee
which gives
\bea &D = 1:&  R_0(\hz) = i \frac{\sqrt{\pi}}{E_c \sqrt{s}} \,
w(-\sqrt{s})\, , \nonumber \\
&D = 2:&  R_0(\hz) = -\frac{e^{-s}}{E_c} E_1(-s) \, ,\\
&D = 3:&  R_0(\hz) = \frac{2}{E_c} \left(i\sqrt{\pi s}
\,w(-\sqrt{s}) - 1\right)\, . \nonumber \eea
In addition to the complex error function \Eq{ComplError}, the
exponential integral $E_1(z) = \int_z^\infty dt \,e^{-t}/t$
enters.

\section{Result for the exact moments}

We list here explicit values for the first exact moments
\Eq{MomH}. For Gauss correlation, we take advantage of the closed
form
\be K_n(\k=0) = E_c^n \frac{\Gamma(D/2+n)}{\Gamma(D/2)} \ee
in $D$ dimensions, and obtain
\bea M_3 &=& \sigma^2 E_c \frac{D}{2} \, ,\\
M_4(\k) &=& \sigma^2 E_c^2 \left[\frac{(D+2)D}{4} + 2(k \ell
)^2\right] \, + \, 3 \,\sigma^4 \, ,\nonumber \\
M_5(\k) &=& \sigma^2 E_c^3 \frac{D+2}{2} \left[
\frac{(D+4)D}{4}+6(k \ell)^2\right]\, + \, 5D \sigma^4 E_c\,
.\nonumber \eea

In the exponential case
\be M_3 = \sigma^2 E_c \ee
holds independent on spatial dimension. The last finite moment is
here
\be \label{M4E} M_4(\k) = \sigma^2 E_c^2 \left[5 +
\frac{4}{3}\left(k \xi \right)^2\right] \, + \, 3 \,\sigma^4 \ee
for $D = 3$. This unexpected termination of the moment expansion
can be understood quite easily: The spectral function
\Eq{SpecDef} decays at large positive energies as
\be A_{\k=0}(\omega) \Rightarrow \sum_\k \frac{\sigma^2
g_\k}{\ek^2} \pi \delta(\hom - \ek) \ee
which follows from plain perturbation theory using \Eq{Self2B} in
leading order. For the Gauss correlation \Eq{GaussCorr},
multiplication with any power of $\hom = \ek$ leads to a convergent
frequency (better momentum) integral. For the exponential case,
however, we have $g_{\k} \propto k^{-(2D+2)}$, and the integrand of
the $n$th moment behaves as $k^{(2n - D - 7)}$. Therefore, the
moments are finite up to $n=3$ ($D=1, 2$) or $n=4$ ($D=3$).

For the sake of completeness, we quote the moments in the
uncorrelated (tight-binding) case as well. $M_0 = 1$, $M_1=0$,
and $M_2=\sigma^2$ hold as before. We write the (simple cubic)
dispersion as
\be \ek = 2TD\left( 1-C(\k)\right)\, , \;\; C(\k) =
\frac{1}{D}\sum_{j=1}^D \cos(k_j\Delta)\ee
and obtain instead of Eqs.\,(\ref{M3})-(\ref{M5})
\bea \label{TBmom} M_3(\k) &=& 2TD\sigma^2 C(\k) \, , \nonumber\\
M_4(\k) &=& (2TD)^2\sigma^2\left(C^2(\k) + \frac{1}{2D}\right)
    + 3 \sigma^4  \, , \\
M_5(\k) &=&
C(\k)\left\{(2TD)^3\sigma^2\left(C^2(\k)+\frac{3}{2D}\right) + 16
TD\sigma^4\right\}\, . \nonumber \eea
%



\begin{thebibliography}{99}
%
\bibitem{Soven}P. Soven, Phys. Rev. \textbf{156}, 809 (1967).
%
\bibitem{VKE}B. Velicky, S. Kirkpatrick, and H. Ehrenreich, Phys.
Rev. \textbf{175}, 747 (1968).
%
\bibitem{Velicky}B. Velicky, Phys. Rev. \textbf{184}, 614 (1969).
%
\bibitem{Cluster}Yu-Tang Shen and Ch. W. Myles, Phys. Rev. B \textbf{30}, 3283 (1984).
%
\bibitem{Eggarter}T. P. Eggarter and A. Troper, Phys. Status Solidi B \textbf{140}, 127 (1987).
%
\bibitem{Mills}R. Mills and P. Ratanavararaksa, Phys. Rev. B \textbf{18}, 5291 (1978);
R. Mills, L. J. Gray, and Th. Kaplan, Phys. Rev. B \textbf{27}, 3252
(1983).
%
\bibitem{Mookerjee}A. Mookerjee and R. Prasad, Phys. Rev. B
\textbf{48}, 17724 (1993).
%
\bibitem{Jarrell}M. Jarrell and H. R. Krishnamurthy, Phys. Rev. B \textbf{63}, 125102 (2001).
%
\bibitem{Maier}Th. Maier, M. Jarrell, Th. Pruscheke, and M. H. Hettler,
Rev. Mod. Phys. \textbf{77}, 1027 (2005).
%
\bibitem{Rowlands}D. A. Rowlands, J. Phys.: Condens. Matter
\textbf{18}, 3179 (2006).
%
\bibitem{Laad}M.S. Laad and L. Craco, J. Phys.: Condens. Matter
\textbf{17}, 4765 (2005).
%
\bibitem{Alvermann}A. Wei\ss{}e, G. Wellein,
A. Alvermann, and H. Fehske, Rev. Modern Physics \textbf{78}, 275
(2006).
%
\bibitem{Baranovski}S. D. Baranovskii and A. L. Efros, Sov. Phys.
Semicond., \textbf{12}, 1328 (1978).
%
\bibitem{Zim89}R. Zimmermann, J. Crystal Growth \textbf{101}, 346 (1990).
%
\bibitem{LeeBajaj}S. M. Lee and K. K. Bajaj, Appl. Phys. Lett.
\textbf{60}, 853 (1992).
%
\bibitem{Lyo}S. K. Lyo, Phys. Rev. B \textbf{48}, 2152 (1993).
%
\bibitem{Kanehisa}M. A. Kanehisa and R. J. Elliott, Phys. Rev. B
\textbf{35}, 2228 (1987).
%
\bibitem{Huber}D. L. Huber and W. Y. Ching, Phys. Rev. B \textbf{39},
8652 (1989).
%
\bibitem{Reineker}P. Reineker, J. K\"ohler, and A. M. Jayannavar,
J. Lumin. \textbf{45}, 102 (1990).
%
\bibitem{Adame}F. Dominguez-Adame, Phys. Rev. B \textbf{51},
12801 (1995).
%
\bibitem{Bakalis}L. D. Bakalis, I. Rubtsov, and J. Knoester, J.
Chem. Physics \textbf{117}, 5393 (2002).
%
\bibitem{GlutschTime}S. Glutsch, D. S. Chemla, and F. Bechstedt
Phys. Rev. B \textbf{54}, 11592 (1996).
%
\bibitem{TakaRev}R. Zimmermann, E. Runge, and V. Savona: Theory of resonant
secondary emission: Rayleigh scattering versus luminescence In:
\textit{Quantum Coherence, Correlation and Decoherence in
Semiconductor Nanostructures} (T. Takagahara ed.), p. 89-165,
Elsevier Science (USA), 2003.
%
\bibitem{SavonaZim}V. Savona and R. Zimmermann,
Phys. Rev. B \textbf{60}, 4928 (1999).
%
\bibitem{Lienau}Ch. Lienau, F. Intonti, T. Guenther, Th. Elsaesser, V. Savona, R.
Zimmermann, and E. Runge, Phys. Rev. B \textbf{69}, 085302 (2004).
%
\bibitem{RungeSSP}E. Runge, Solid State Physics Vol. \textbf{57},
(H. Ehrenreich and F. Saepen ed.), p. 149-305, Academic Press, San
Diego, 2002.
%
\bibitem{Grochol}M. Grochol, F. Grosse, and R. Zimmermann,
Phys. Rev. B \textbf{71}, 125339 (2005).
%
\bibitem{Kraich}R. H. Kraichnan, Phys. Rev. \textbf{109}, 1407 (1958).
%
\bibitem{Stroucken}T. Stroucken, C. Anthony, A. Knorr, P. Thomas,
and S. W. Koch, Phys. Status Solidi B \textbf{188}, 539 (1995).
%
\bibitem{GlutschMom}St. Glutsch and F. Bechstedt, Phys. Rev. B \textbf{50},
7733 (1994).
%
\bibitem{Vogt}M. Vogt, R. Zimmermann, and R. J. Needs, Phys. Rev.
B \textbf{69}, 045113 (2004).
%
\bibitem{OptFluc}I. M. Lifshits, S. A. Gredeskul, and L. A. Pastur,
\textit{Introduction to the Theory of Disordered systems}, Wiley,
New York, 1988.
%
\bibitem{Mills84}A. K. Sen, R. Mills, Th. Kaplan, and L. J. Gray, Phys. Rev. B \textbf{30}, 5686 (1984).
%
\end{thebibliography}
\end{document}